\documentclass[preprint,showpacs,preprintnumbers,amsmath,amssymb]{revtex4}

\usepackage{color,changes}
\usepackage{soul}
\usepackage{hyperref}

\definecolor{darkpastelgreen}{rgb}{0.01, 0.75, 0.24}

\newcommand{\beq}{\begin{equation}}
\newcommand{\eeq}{\end{equation}}
\newcommand{\bea}{\begin{eqnarray}}
\newcommand{\eea}{\end{eqnarray}}
\def\l{\left}
\def\r{\right}

\newcommand{\eq}[1]{Eq.~(\ref{#1})}

\begin{document}


\title{Addendum to ``Invisible Higgs decay width versus dark matter direct detection cross section in Higgs portal dark matter models" }


\author{Seungwon Baek}
\email[]{sbaek1560@gmail.com}
\affiliation{Department of Physics, Korea University, Seoul 02841, South Korea}
\author{P. Ko}
\email[]{pko@kias.re.kr}
\affiliation{School of Physics, KIAS, Seoul 02455, South Korea}
\author{Wan-Il Park}
\email[]{wipark@jbnu.ac.kr}
\affiliation{Division of Science Education and Institute of Fusion Science, Jeonbuk National University, Jeonju 54896, South Korea}

\date{\today}

\begin{abstract}
This article is an addendum to Ref.~\cite{Baek:2014jga}.
Here, we discuss the invisible Higgs decay width 
$\Gamma_{h}^{\rm inv}$ in the Higgs portal vector dark matter (VDM) model in the limit 
$m_V \rightarrow 0^+$.  In the effective field theory (EFT) approach where the VDM mass is 
attributed to the St\"{u}ckelberg mechanism, $( \Gamma_{h}^{\rm inv} )_{\rm EFT}$ is divergent,  
which is unphysical and puzzling. 
On the other hand $( \Gamma_{h}^{\rm inv} )_{\rm UV}$ becomes finite in a UV completion, 
where the VDM mass is generated by the dark Higgs mechanism. Then we can take  the limit 
$m_V \rightarrow 0^+$ by taking either {\it (i)}  the dark  gauge coupling $g_X \rightarrow 0^+$ with 
a fixed dark Higgs vacuum expectation value $v_\Phi$, or  {\it (ii)} $v_\Phi \to 0^+$ with a fixed $g_X$.
Such a difference in the behavior of $\Gamma_{h}^{\rm inv}$ in the massless VDM limit 
demonstrates another limitation of EFT for the Higgs portal VDM, and the importance of gauge-invariant 
and renormalizable models for the Higgs portal  VDM.
\end{abstract}


\maketitle

\section{Invisible decay width of the Higgs boson}

Higgs portal (HP) interactions are simple and generic nongravitational channels for communication between 
the visible and the dark sectors.  
Quite often the effective field theory (EFT) approach is taken for concrete realizations of  HP interactions
(see Ref. \cite{Arcadi:2020jqf} for a recent review on this approach).
However, some aspects of HP 
interactions in the EFT approach appear to be qualitatively 
different from those in UV-completions with gauge-invariance, unitarity and renormalizability.  
These issues were discussed in detail in Refs. \cite{Baek:2011aa,Baek:2012uj} for Higgs portal singlet fermion dark matter (HP SFDM), 
and Refs. \cite{Farzan:2012hh,Baek:2012se,Ko:2014gha} for HP VDM. 
In particular, the correlations between the invisible Higgs decay 
branching ratio and the spin-independent DM-nucleus scattering cross section for direct detection
in the Higgs portal DM models, both in the EFT and in UV completions were discussed
in Ref. \cite{Baek:2014jga}.  
Various collider signatures of gauge-invariant renormalizable HP DM models were discussed in 
Refs. \cite{Ko:2016xwd,Kamon:2017yfx,Dutta:2017sod,Ko:2018mew}, 
including the discussions on how unitarity is recovered at high-energy colliders contrarily to the EFT approaches. 

One of the most pronounced differences between the EFT and the full UV completions 
is \textit{the invisible decay width of the standard model Higgs to VDM ($\Gamma_h^{\rm inv}$)} 
in the Higgs portal VDM scenarios.  
In the massless VDM limit, the invisible Higgs decay width diverges in the EFT approach, whereas it remains finite in UV completions, as described below.
Such an appearance of a physically unacceptable observable indicates another limitation of the EFT approach, demanding care in the interpretation of the results of the EFT approach. 
In this addendum, we will demonstrate that this divergent catastrophe  
in the Higgs invisible decay width for the Higgs portal VDM in the EFT approach 
disappears in gauge-invariant renormalizable UV completions discussed in Refs. \cite{Baek:2012se,Ko:2014gha}.  
The analytic expression for $( \Gamma_h^{\rm inv} )_{\rm UV}$ in a UV completion was given in Ref. \cite{Baek:2014jga} already. But its behavior in the limit of $m_V \rightarrow 0^+$ was not discussed 
in detail therein.  The aim of this addendum is to complete discussions on what happens to 
$\Gamma_h^{\rm inv}$ when we take $m_V \rightarrow 0^+$ limit. 

\section{EFT prediction}
In the EFT approach for the Higgs portal VDM, the model Lagrangian  
is defined as \cite{Lebedev:2011iq} 
\begin{equation} \label{eq:lag_eff}
{\cal L}_{\rm VDM} = -\frac{1}{4} V_{\mu\nu} V^{\mu\nu} + \frac{1}{2} m_0^2 V_\mu V^\mu 
+ \frac{1}{2} \lambda_{VH} V_\mu V^\mu H^\dagger H - \frac{\lambda_V}{4\!} (V_\mu V^\mu)^2 ,
\end{equation}
assuming that VDM is the gauge boson for dark $U(1)$ gauge group. 
The VDM mass is supposed to be generated by the St\"{u}ckelberg mechanism. 
Then the DM phenomenology can be described in terms of two parameters, 
$ m_V$ and $\lambda_{VH}$. 

One of the important physical observables in the Higgs portal DM models (and in general) is the invisible Higgs decay width. 
In the EFT approach for HP VDM, this quantity is given by~\cite{Lebedev:2011iq} 
\beq \label{eq:inv-decay-eff}
( \Gamma_h^{\rm inv} )_{\rm EFT} 
= \frac{\lambda_{VH}^2}{128 \pi} \frac{v_H^2 m_h^3}{m_V^4} 
\l( 1 - \frac{4 m_V^2}{m_h^2} + 12 \frac{m_V^4}{m_h^4} \r) \l( 1 - \frac{4 m_V^2}{m_h^2} \r)^{1/2} ,
\eeq
where $m_V$ is the mass of the VDM, and $m_h$ and $v_H = \sqrt{2} \langle H^0 \rangle \approx 246~{\rm GeV}$
are  the mass and the vacuum expectation value of the canonically normalized neutral component of Higgs field, respectively.
Within the EFT approach, $m_V^2 = m_0^2 + \lambda_{VH} v_H^2/2$ with $m_0$ and $\lambda_{VH}$ 
being arbitrary free parameters subject to only phenomenological constraints.
Hence, there is no definite relation between $m_V$ and $\lambda_{VH}$ unless $m_0 =0$ 
which is however already ruled out by constraints from the XENON1T experiment 
(for example, see Fig. 1 in Ref. \cite{Arcadi:2020jqf}).
Note that the prefactor in (\ref{eq:inv-decay-eff}) becomes 
independent of $m_V$  for $m_0=0$.  Then the Higgs decay width is predicted to
be $m_h^3/32\pi v_H^2\approx 320$~MeV in the limit $m_V\to 0^+$, which is already well above 
the current LHC upper bound of 14~MeV at 95\% C.L.~\cite{Aaboud:2018puo}.
Therefore, we should take $m_0 \neq 0$, and 
the invisible decay width $( \Gamma_h^{\rm inv})_{\rm EFT}$ will grow indefinitely 
when $m_V \rightarrow 0^+$,  
which is the well-known puzzle in the Higgs portal VDM in the EFT approach.

\section{Renormalizable and gauge-invariant theory}

The model Lagrangian in Eq. (1) violates gauge invariance, unitarity and renormalizability,  and has to be fixed.  
One simple UV completion is to consider the Abelian-Higgs model in the dark sector, and the VDM mass 
is generated by dark Higgs mechanism  \cite{Farzan:2012hh,Baek:2012se,Ko:2014gha}. 
The SM and the dark sectors will communicate through the Higgs portal coupling 
($\lambda_{H \Phi}$ in Eq. (3) below) between the SM Higgs doublet $H$ and the dark Higgs $\Phi$,
both of which develop nonzero vacuum expectation values. 
The renormalizable and gauge-invariant model Lagrangian is simply given by \cite{Farzan:2012hh,Baek:2012se,Ko:2014gha,Baek:2014jga}
\begin{eqnarray}
{\cal L} &=& - \frac{1}{4} V_{\mu\nu} V^{\mu\nu} + D_\mu \Phi^\dagger D^\mu \Phi 
 - \frac{\lambda_\Phi}{4} \left( \Phi^\dagger \Phi - \frac{v_\Phi^2}{2} \right)^2 
\nonumber \\
&&- \lambda_{H\Phi} \left(H^\dagger H -  \frac{v_H^2}{2} \right) 
\left( \Phi^\dagger \Phi - \frac{v_\Phi^2}{2} \right) 
- \frac{\lambda_H}{4} 
\left( H^\dagger H - \frac{v_H^2}{2} \right)^2 ,
\label{eq:lag_full} 
\end{eqnarray}
where $D_\mu = \partial_\mu + i g_X Q_\Phi V_\mu$ with $Q_\Phi$ being the dark charge of $\Phi$.
There are two neutral scalar bosons; a 125 GeV SM Higgs-like one, and the other being 
mostly a singlet-scalar dark Higgs boson.  The dark Higgs boson can play interesting and 
important roles in DM phenomenology,  particle physics, and  cosmology, including 
such as Higgs-portal assisted Higgs inflation \cite{Ko:2014eia} (see Refs. \cite{KO:2016gxk,Ko:2018qxz} 
for reviews).

After electroweak and dark-gauge symmetry breaking, the SM Higgs boson and the dark Higgs boson 
will be mixed with each other through the Higgs-portal interaction term, $\lambda_{H\Phi}$.
In the unitary gauge, we have 
\begin{equation}
H(x)  =  \left( \begin{array}{c} 0 \\ \frac{1}{\sqrt{2}} \left( v_H + \tilde{h} (x) \right)  \end{array}  \right) , \ \ \
\Phi (x)  =  \frac{1}{\sqrt{2}} ( v_\Phi + \varphi (x) )   
\end{equation}
The mass matrix in the basis $( \tilde{h} ,\varphi)$ can be written as
\begin{equation}
\left( \begin{array}{cc} 
\lambda_H v_H^2/2 & \lambda_{H\Phi} v_H v_\Phi \\
\lambda_{H\Phi} v_H v_\Phi &  \lambda_\Phi v_\Phi^2/2 \end{array} \right) =
\left( \begin{array}{cc}
m_1^2 c_\alpha^2 + m_2^2 s_\alpha^2 & (m_2^2 - m_1^2 ) s_\alpha c_\alpha 
\\
 (m_2^2 - m_1^2 ) s_\alpha c_\alpha & m_1^2 s_\alpha^2 + m_2^2 c_\alpha^2 
 \end{array} \right).
\end{equation}
The mass eigenstates $( H_1, H_2 )$ are defined by the following relation:
\begin{equation}
\left( \begin{array}{c}  H_1 \\ H_2  \end{array}  \right)  
= \left( \begin{array}{cc} 
           \cos\alpha & -\sin\alpha \\
           \sin\alpha & \cos\alpha 
           \end{array} \right)
\left( \begin{array}{c}  \tilde{h} (x)  \\ \varphi(x)  \end{array}  \right),  
\end{equation}
where $H_1 \equiv h$ is identified as the 125 GeV Higgs boson observed at the LHC, and the mixing angle $\alpha$ between $\tilde{h}$ and $\varphi$ is given by 
\begin{equation}
\tan 2 \alpha = \frac{4 \lambda_{H\Phi} v_H v_\Phi}{\lambda_{\Phi} v_\Phi^2-\lambda_{H} v_H^2}.
\label{eq:alpha} 
\end{equation}

The invisible Higgs decay width within the full gauge-invariant and renormalizable model (3) is 
given by \cite{Baek:2014jga}
\beq 
\l( \Gamma_h^{\rm inv} \r)_{\rm UV}= \frac{g_X^2 Q_\Phi^2}{32 \pi} \frac{m_h^3}{m_V^2} \sin^2 \alpha 
\l( 1 - \frac{4 m_V^2}{m_h^2} + 12 \frac{m_V^4}{m_h^4} \r) \l( 1 - \frac{4 m_V^2}{m_h^2} \r)^{1/2}~ . 
\label{eq:inv_full} 
\eeq
%
In the limit $m_V \rightarrow 0^+$, the main contribution to (\ref{eq:inv_full}) comes from the
longitudinally-polarized $V$s, where the polarization vector is in the form $\epsilon_\mu(k) \approx k_\mu/m_V$.
This also explains the enhancement factor $m_h^2/m_V^2$ in Eq.~(\ref{eq:inv_full}).
The invisible Higgs decay width is constrained by the signal strengths of Higgs boson in various 
production and decay channels, and the upper limits on the Higgs invisible branching ratio as well as  
on the nonstandard Higgs decay width (see, for example, \cite{Chpoi:2013wga,Cheung:2015dta}).

The critical difference of \eq{eq:inv_full} compared with the EFT result in \eq{eq:inv-decay-eff} is that 
$m_V^2 = g_X^2 Q_\Phi^2 v_\Phi^2$ in the UV completed model.  Note that the massless VDM limit,
$m_V \to 0^+$, can be achieved by taking either $g_X Q_\Phi \to 0^+$ or $v_\Phi \to 0^+$
in \eq{eq:inv_full}. We find that in both cases the Higgs invisible decay widths are finite, and physically 
sensible results are obtained as described below.

\subsection{$g_X Q_\Phi \to 0^+$ with $v_\Phi \neq 0$ fixed}
For a finite fixed $v_\Phi$, we notice that the mixing angle $\alpha$ is fixed and finite, 
since the $2 \times 2$ scalar-mass matrix in Eq. (5) is independent of $g_X$. 
And the prefactor in Eq. (\ref{eq:inv_full}) becomes
\[
\frac{g_X^2 Q_\Phi^2}{m_V^2} = \frac{g_X^2 Q_\Phi^2}{ g_X^2 Q_\Phi^2 v_\Phi^2} = \frac{1}{v_\Phi^2} = {\rm finite}.
\]
Then  one obtains
%
\beq
\l( \Gamma_h^{\rm inv} \r)_{\rm UV} =
\frac{1}{32 \pi} \frac{m_h^3}{v_\Phi^2} \sin^2 \alpha ,
\label{eq:inv_full_limit1} 
\eeq
which is finite irrespective of the VDM mass and is physically sensible \footnote{This behavior of $( \Gamma_h^{\rm inv})_{\rm UV}$ in the limit {\bf A} was presented by one of the authors (P.K.) in two DM@LHC workshops, one at Amsterdam (2016) and the other at Irvine (2017).}.
Note that, $m_h \gg m_V$ in this limit and the VDMs produced in the decay of the SM Higgs are highly boosted.
Hence, the decay rate in \eq{eq:inv_full_limit1} is actually mostly from the longitudinal mode of the VDM.
Then, it is clear that from the Goldstone boson equivalence theorem one should have the same rate as the one in \eq{eq:inv_full_limit1} for the decay of the SM Higgs to its associated Goldstone bosons when $g_X Q_\Phi = 0$.

Indeed, for $g_X Q_\Phi \equiv 0$ and $v_\Phi \neq 0$, there is no interaction between $V_\mu$ and the dark Higgs $\Phi$.
Specifically the Higgs-V-V interaction vanishes identically- 
\[
 -g_X^2 Q_\Phi^2 v_\Phi \sin\alpha V_\mu V^\mu h \equiv 0 ,
\]
and consequently  the partial width  $\Gamma(h \to VV)$ vanishes.
%
%
Since $V$ is massless for $g_X Q_\Phi = 0$, the Goldstone boson $a_\Phi$ from $\Phi$ is not absorbed into the longitudinal component of $V$ but becomes a physical degree of freedom. That is, the dark $U(1)$ symmetry acts as a global symmetry.
In this case the Higgs boson $h$ can decay into a pair of the Goldstone bosons through the mixing 
with the dark Higgs boson, and the partial decay width is found to be~\cite{Baek:2013ywa},
\begin{eqnarray}
\Gamma(h \to a_\Phi a_\Phi) = \frac{ \sin^2\alpha \, m_h^3}{32 \pi v_\Phi^2} ,
\end{eqnarray}
which is exactly what we obtain from \eq{eq:inv_full} with $g_X Q_\phi \to 0^+$ 
as shown in \eq{eq:inv_full_limit1}.


\subsection{$v_\Phi \to 0^+$ with $g_X Q_\Phi$ fixed}
Another possibility for a massless VDM would be taking $v_\Phi \rightarrow 0^+$ with a finite value of 
$g_X$.    In this limit, the mixing angle $\alpha$  defined in Eq.~(\ref{eq:alpha})  is approximated as
\begin{equation}
\alpha \xrightarrow{v_\Phi \to 0^+} -\frac{2 \lambda_{H \Phi} v_\Phi}{\lambda_H v_H} . 
\end{equation}
Then the prefactor (including the mixing factor) in $( \Gamma_h^{\rm inv} )_{\rm UV}$ [Eq. (8)] becomes 
\begin{equation}
\frac{g_X^2 Q_\Phi^2}{m_V^2} \sin^2 \alpha \xrightarrow{v_\Phi \to 0^+}{\frac{4 \lambda_{H \Phi}^2}{\lambda_H^2 v_H^2}} = \frac{2 \lambda_{H \Phi}^2}{\lambda_H m_h^2} = {\rm finite} ,
\end{equation}
where  in the second equality we have used $m_h^2 \to \lambda_H v_H^2/2$ as 
$v_\Phi \to 0^+$.
Then the invisible Higgs decay rate in \eq{eq:inv_full} can be approximated as
\beq
\l( \Gamma_h^{\rm inv} \r)_{\rm UV} \xrightarrow{v_\Phi \to 0^+} \frac{1}{16 \pi} \frac{\lambda_{H \Phi}^2 m_h}{\lambda_H} \ , 
\label{eq:inv_full_limit2} 
\eeq
which is again finite.  
Note that \eq{eq:inv_full_limit2} is exactly what one finds for the decay of the SM-like Higgs to Goldstone bosons in the linear representation of $\Phi$ in the broken phase.
Hence, we find that in the broken phase (i.e., $v_\Phi \neq 0$)
whichever limit we take to get a massless VDM limit,  namely either $g_X Q_\Phi \to 0^+$ or 
$v_\Phi \to 0^+$ to realize $m_V \to 0^+$, the invisible decay rate of the SM Higgs in the UV complete model is finite and physically consistent with the expectation from the Goldstone boson equivalence theorem, as opposed to the case of the EFT approach discussed in Sec. II. 
%
%

\subsection{Unbroken $U(1)$ case with $g_X Q_\Phi \neq 0$ and $m_V = 0$ }
For completeness, we briefly discuss the unbroken $U(1)$ case with $g_X Q_\Phi \neq 0$,  
for which the dark $U(1)$ gauge boson remains massless, $m_V \equiv 0$.  
In this case,  we have
$\Gamma(h \to VV)\equiv 0$.
The scalar potential form of Eq.~(\ref{eq:lag_full})  in the broken phase for the $U(1)$ 
is not adequate here.   
Instead we shall write the scalar potential as
\beq
{\cal L} \supset - {1 \over 2} m_\Phi^2 \Phi^\dagger \Phi - {\lambda_\Phi \over 4} \l(\Phi^\dagger \Phi\r)^2 -\lambda_{H\Phi} \left( H^\dagger H - \frac{v_H^2}{2} \right) \Phi^\dagger \Phi 
- \frac{\lambda_H}{4} \left( H^\dagger H - \frac{v_H^2}{2} \right)^2,
\eeq
where $m_\Phi(>0)$ is the physical mass of $\Phi$.
Since $\Phi$ is the lightest (actually the only) charged field under the  
unbroken dark $U(1)$ gauge symmetry, it is stable and makes
complex scalar dark matter.
The massless photon $V$ plays the role of the dark photon 
and can contribute to the energy budget of the Universe as a part of dark 
radiation (DR). 
Although it has a cold dark matter candidate $\Phi$, the model is 
no long the same as our original Higgs portal VDM.   
Now it is a Higgs portal complex scalar DM model.
The contribution to $\l( \Gamma_h^{\rm inv} \r)_{\rm UV} $ comes from $h \to \Phi \Phi^\dagger$, 
if kinematically allowed, and the result is 
\bea
\l( \Gamma_h^{\rm inv} \r)_{\rm UV}  = \frac{1}{8\pi} \frac{ \lambda_{H\Phi}^2 m_h}{\lambda_H}\l(1 - {4 m_\Phi^2 \over m_h^2}\r)^{1/2} ,
\label{eq:inv_v0_limit}
\eea
which is clearly finite.   
Note that the rate in \eq{eq:inv_full_limit2} in the limit of $m_V \to 0$ differs from the one in \eq{eq:inv_v0_limit} in the limit of $m_\Phi \to 0$ only by 
a factor of $2$.
This can be understood from the fact that in the linear representation of $\Phi$ in both  broken and 
unbroken phases the three-point couplings of $h$ to the real components and  to the imaginary components of $\Phi$ are equal to each other.
Hence, if kinematically allowed, the decay rate of $h \to \Phi \Phi^\dag$ should be larger than 
that for the decay to Goldstone bosons by a factor of 2, modulo the phase space factor.  

\section{Conclusion}
In conclusion, the invisible decay rate of the SM Higgs $\l( \Gamma_h^{\rm inv} \r)_{\rm UV}$ in the UV-complete model of \eq{eq:lag_full} remains finite and physically sensible in the $m_V\rightarrow 0^+$ limit obtained by taking either $g_X Q_\Phi \to 0^+$ or $v_\Phi \to 0^+$, which is in sharp contrast with Eq. (\ref{eq:inv-decay-eff}) derived in the EFT approach.
Also note that the cases of $g_X Q_\Phi \equiv 0$ or $v_\Phi \equiv 0$ are different from the VDM 
we are discussing as described above,  since in such cases the vector field is massless and becomes 
dark radiation instead of being dark matter. 

Before closing, let us make additional comments on what happens in another possible UV completion. 
It is well known that there are, in general,  more than one UV completion for a given EFT operator.  
As an alternative to the UV completion [\eq{eq:lag_full}] for the Higgs portal VDM, 
a radiatively generated Higgs-portal VDM model was discussed in Ref. \cite{DiFranzo:2015nli}, 
where  new heavy fermions were introduced in \eq{eq:lag_full} in order to generate 
(radiatively) the Higgs portal interaction term in \eq{eq:lag_eff}. 
It is found that the one-loop induced invisible decay rate of Higgs is also finite in this scenario.  
It is another example that the Higgs invisible decay width is finite in gauge-invariant and 
renormalizable UV completions for Higgs portal VDM.

\begin{acknowledgments}
We are grateful to P. J. Fox for raising this issue in the EFT approach. 
The work is supported in part by KIAS Individual Grant No. PG021403 (P.K.) at Korea Institute for Advanced Study, by National Research Foundation of Korea Grant No. NRF-2018R1A2A3075605 (S.B.) and No. NRF-2019R1A2C3005009 (P.K.), funded by the Korea government (Ministry of Science and ICT), and by Basic Science Research Program through the National Research Foundation of Korea funded by the Ministry of Education Grant No. 2017R1D1A1B06035959 (W.I.P.).
\end{acknowledgments}


\end{document}